\documentclass[aps,prl,preprint,tightenlines,superscriptaddress,showpacs,byrevtex]{revtex4}
\usepackage{graphicx}
\usepackage{dcolumn}
\usepackage{bm}

\begin{document}


\preprint{\vbox{ \hbox{   }
                \hbox{BELLE-CONF-0201}
                \hbox{Parallel Session: 7}
                \hbox{ABS688}
}}

\def\st{\scriptstyle}
\def\sst{\scriptscriptstyle}
\def\mco{\multicolumn}
\def\epp{\epsilon^{\prime}}
\def\vep{\varepsilon}
\def\ra{\rightarrow}
\def\ppg{\pi^+\pi^-\gamma}
\def\vp{{\bf p}}
\def\ko{K^0}
\def\kb{\bar{K^0}}
\def\al{\alpha}
\def\ab{\bar{\alpha}}
\def\be{\begin{equation}}
\def\ee{\end{equation}}
\def\bea{\begin{eqnarray}}
\def\eea{\end{eqnarray}}
\def\CPbar{\hbox{{\rm CP}\hskip-1.80em{/}}}

\def\tbzresultmean{1.554}
\def\tbzresultstat{\pm 0.030}
\def\tbzresultsyst{\pm0.019}

\def\tbmresultmean{1.695}
\def\tbmresultstat{\pm 0.026}
\def\tbmresultsyst{\pm 0.015}

\def\rbmresultmean{1.091}
\def\rbmresultstat{\pm 0.023}
\def\rbmresultsyst{\pm 0.014}

\def\yresultmean{0.114}
\def\yresultstat{{_{-0.064}^{+0.060}}}
\def\yresultsyst{{_{-0.011}^{+0.012}}}

\def\yinterval{-0.013 < y_{CP} < 0.234}

\def\tbzresult{\tbzresultmean \tbzresultstat\;\mbox{(stat)} \; \tbzresultsyst\;\mbox{(syst)}}
\def\tbmresult{\tbmresultmean \tbmresultstat\;\mbox{(stat)} \tbmresultsyst\;\mbox{(syst)}}
\def\rbmresult{\rbmresultmean \rbmresultstat\;\mbox{(stat)} \rbmresultsyst\;\mbox{(syst)}}
\def\yresult{\yresultmean \; \yresultstat \; \yresultsyst}


\def\rtau{r_\tau}

\def\yllimit{-0.36}
\def\yulimit{0.35}
\def\tbzdstlnu{1.50\pm0.06^{+0.06}_{-0.04}}
\def\tbmdstlnu{1.54\pm0.10^{+0.14}_{-0.07}}
\def\tbzdstpm{1.55^{+0.18}_{-0.17}{}^{+0.10}_{-0.07}}
\def\tbzdppm{1.41^{+0.13}_{-0.12}\pm0.07}
\def\tbmdzpm{1.73\pm0.10\pm0.09}
\def\tbmpsikm{1.87^{+0.13}_{-0.12}{}^{+0.07}_{-0.14}}
\def\tbzpsiks{1.54^{+0.28}_{-0.24}{}^{+0.11}_{-0.19}}
\def\tbzpsikst{1.56^{+0.22}_{-0.19}{}^{+0.09}_{-0.15}}
\def\Tbzdstlnu{$\taubz=(\tbzdstlnu)$~ps}
\def\Tbmdstlnu{$\taubm=(\tbmdstlnu)$~ps}
\def\Tbzdstpm{$\taubz=(\tbzdstpm)$~ps}
\def\Tbzdppm{$\taubz=(\tbzdppm)$~ps}
\def\Tbmdzpm{$\taubm=(\tbmdzpm)$~ps}
\def\Tbmpsikm{$\taubm=(\tbmpsikm)$~ps}
\def\Tbzpsiks{$\taubz=(\tbzpsiks)$~ps}
\def\Tbzpsikst{$\taubz=(\tbzpsikst)$~ps}
\def\Tbzresult{$\taubz=(\tbzresult)$~ps}
\def\Tbmresult{$\taubm=(\tbmresult)$~ps}
\def\Rbmresult{$\rbm=\rbmresult$}
\def\Yresult{$\ycp=\yresult$}
\def\Ylimit{$\yllimit<\ycp<\yulimit$}
\def\intl{5.1}


\newcommand{\nim}[3]{Nucl. Inst. and Meth. {\bf #1} #2 (#3)}
\newcommand{\prld}[3]{Phys. Rev. Lett. {\bf #1} #2 (#3)}
\newcommand{\prdd}[3]{Phys. Rev. D {\bf #1} #2 (#3)}
\newcommand{\prxd}[3]{Phys. Rev. {\bf #1} #2 (#3)}
\newcommand{\plb}[3]{Phys. Lett. B {\bf #1} #2 (#3)}
\newcommand{\npa}[3]{Nucl. Phys. A {\bf #1} #2 (#3)}
\newcommand{\apphyslet}[3]{App. Phys. Lett. {\bf #1} #2 (#3)}
\newcommand{\zphysa}[3]{Z. Phys. A. {\bf #1} #2 (#3)}
\newcommand{\zpc}[3]{Z. Phys. C {\bf #1} #2 (#3)}
\newcommand{\jphys}[3]{J. Phys. {\bf #1} #2 (#3)}

\def\etal{{\it et al.}}
\def\chisqndf{{\chi}^2/n}
\def\chisq{\chi^2}
\def\Chisqndf{$\chisqndf$}
\def\Chisq{$\chisq$}
\def\NDF{$N.D.F.$}
\def\degree{{}^{\circ}}
\def\dG{\Delta\Gamma}
\def\dM{\Delta M}
\def\rmix{R_{\rm mix}}
\def\amix{A_{\rm mix}}
\def\re{{\cal R}e}
\def\im{{\cal I}m}
\def\bzb{{\overline{B}{}^0}}
\def\bb{{\overline{B}{}^0}}
\def\kz{{K{}^0}}
\def\kzb{{\overline{K}{}^0}}
\def\kb{{\overline{K}{}^0}}
\def\bz{{B^0}}
\def\bh{{B_H}}
\def\bl{{B_L}}
\def\bm{{B^-}}
\def\bp{{B^+}}
\def\taubz{\tau(\bzb)}
\def\taubm{\tau(\bm)}
\def\rbm{\taubm/\taubz}
\def\Bzb{$\bzb$}
\def\Bm{$\bm$}
\def\Fb{fb$^{-1}$}
\def\ycp{y_{CP}}
\def\piz{\pi^0}
\def\pip{\pi^+}
\def\pim{\pi^-}
\def\kz{K^0}
\def\kp{K^+}
\def\km{K^-}
\def\ks{K_S^0}
\def\kl{K_L^0}
\def\kb{\overline{K}}
\def\rhom{\rho^-}
\def\bbar{\overline{B}}
\def\bbbar{B\bbar}
\def\bzbzbar{\bz\bzb}
\def\BBbar{$\bbbar$}
\def\BzBzbar{$\bzbzbar$}
\def\ccbar{c\overline{c}}
\def\CCbar{$\ccbar$}
\def\dstar{D^{*}}
\def\dstarz{{D^{*0}}}
\def\dstarp{D^{*+}}
\def\dstarzb{\overline{D}^{*0}}
\def\dstarm{D^{*-}}
\def\dmdstp{\Delta M_{\dstarp}}
\def\dmdstz{\Delta M_{\dstarz}}
\def\dmdst{\Delta M_{\dstar}}
\def\DM{$\Delta M$}
\def\Dstar{$\dstar$}
\def\Dstarz{$\dstarz$}
\def\Dstarp{$\dstarp$}
\def\Dstarzb{$\dstarzb$}
\def\Dstarm{$\dstarm$}
\def\nub{\overline{\nu}}
\def\jpsi{{J/\psi}}
\def\dz{D^0}
\def\Dz{$\dz$}
\def\Dt{\Delta t}
\def\Dz{\Delta z}
\def\dplus{D^+}
\def\Dp{$\dplus$}
\def\dzb{\overline{D}{}^0}
\def\Dzb{$\dzb$}
\def\dstl{\dstar\ell}
\def\dstpl{\dstarp\ell}
\def\dstzl{\dstarz\ell}
\def\dstml{\dstarm\ell}
\def\dstzbl{\dstarzb\ell}
\def\bdstlnu{\overline{B}\to\dstar\ell^-\nub}
\def\bdstxlnu{\overline{B}\to\dstar X\ell^-\nub}
\def\bzdstlnu{\bzb\to\dstarp\ell^-\nub}
\def\bzdstxlnu{\bzb\to\dstarp X\ell^-\nub}
\def\bmdstlnu{\bm\to\dstarz\ell^-\nub}
\def\bmdstxlnu{\bm\to\dstarz X\ell^-\nub}
\def\bmdstpxlnu{\bm\to\dstarp X\ell^-\nub}
\def\bpsik{\overline{B}\to\jpsi\kb}
\def\bdpi{\bbar\to D\pi}
\def\bzdstpm{\bzb\to\dstarp\pim}
\def\bzdstrhom{\bzb\to\dstarp\rhom}
\def\bzdppm{\bzb\to\dplus\pim}
\def\bmdzpm{\bm\to\dz\pim}
\def\bpsiks{B\to\jpsi\ks}
\def\bzbpsiks{\bzb\to\jpsi\ks}
\def\bzpsiks{\bz\to\jpsi\ks}
\def\bzbpsikst{\bzb\to\jpsi\kstarzb}
\def\bmpsikm{\bm\to\jpsi\km}
\def\nbdstlnu{N_{\bdstlnu}}
\def\nbdstxlnu{N_{\bdstxlnu}}
\def\nbb{N_{\bbbar}}
\def\ncc{N_{\ccbar}}
\def\kpi{K^-\pi^+}
\def\kpipiz{\km\pip\piz}
\def\kpipipi{\km\pip\pip\pim}
\def\kstarzb{\overline{K}{}^{*0}}
\def\Kstarzb{$\kstarzb$}
\def\UPS{$\Upsilon(4S)$}
\def\bgu{(\beta\gamma)_\Upsilon}
\def\pdstl{\mathbf{p}_{\dstar\ell}}
\def\pb{\mathbf{p}_{B}}
\def\Gevc{GeV/$c$}
\def\Gevcsq{GeV/$c^2$}
\def\Mevc{MeV/$c$}
\def\Mevcsq{MeV/$c^2$}
\def\degree{{}^{\rm o}}
\def\Degree{${}^{\rm o}$}
\def\dE{\Delta E}
\def\DE{$\dE$}
\def\mb{M_{\rm bc}}
\def\Mb{$\mb$}
\def\micron{$\mu$m}
\def\delz{\Delta z}
\def\delt{\Delta z}
\def\dt{\Delta t}
\def\dzb{\delz_B}
\def\dtb{\dt_B}
\def\dtp{\dt'}
\def\DTb{$\dtb$}
\def\dtrec{\dt_{rec}}
\def\dtgen{\dt_{gen}}
\def\dzrec{\delz_{rec}}
\def\dzgen{\delz_{gen}}
\def\sigdt{\sigma_{\dt}}
\def\sigpdt{\sigma'_{\dt}}
\def\sigmisdt{\sigma_{tail}^{\dt}}
\def\sigdz{\sigma_{\delz}}
\def\sigtdz{\tilde{\sigma}_{\delz}}
\def\sigtmisdz{\tilde{\sigma}_{tail}^{\delz}}
\def\arec{\alpha_{rec}}
\def\aasc{\alpha_{asc}}
\def\sigzrec{\sigma_{z}^{rec}}
\def\sigzasc{\sigma_{z}^{asc}}
\def\sigtzrec{\tilde{\sigma}_{z}^{rec}}
\def\sigtzasc{\tilde{\sigma}_{z}^{asc}}
\def\sigk{\sigma_{K}}
\def\sigmisk{\sigma_{tail}^{K}}
\def\smis{S_{tail}}
\def\scharm{S_{charm}}
\def\sdet{S_{det}}
\def\sdata{s_{\rm data}}
\def\scmis{S_{tail}^{charm}}
\def\sdmis{S_{tail}^{det}}
\def\smisbg{S_{tail}^{\rm bkg}}
\def\sbg{S_{\rm bkg}}
\def\mudz{\mu_{\delz}}
\def\mumisdz{\mu_{tail}^{\delz}}
\def\mumisbg{\mu_{tail}^{\rm bkg}}
\def\muz{\mu_0}
\def\mumisz{\mu_{tail}^0}
\def\mudt{\mu_{\dt}}
\def\mumisdt{\mu_{tail}^{\dt}}
\def\amu{\alpha_\mu}
\def\amismu{\alpha_{tail}^{\mu}}
\def\mubg{\mu_{\rm bkg}}
\def\fmis{f_{tail}}
\def\fmisbg{f_{tail}^{\rm bkg}}
\def\flmbg{f_{\lambda \rm bkg}}
\def\fsig{f_{\rm sig}}
\def\fbkg{f_{\rm bkg}}
\def\fbg{f_{\rm bkg}}
\def\Fbg{F_{\rm bkg}}
\def\Fsig{F_{\rm sig}}
\def\lmbg{\lambda_{\rm bkg}}
\def\pbg{p_{\rm bkg}}
\def\fm{f_-}
\def\fz{f_0}
\def\ffdst{f_{f \dstar}}
\def\pfdst{p_{f \dstar}}
\def\lfdst{\lambda_{f \dstar}}
\def\ffl{f_{f \ell}}
\def\fbb{f_{\bbbar}}
\def\fcc{f_{\ccbar}}
\def\ftbg{f_{\tau {\rm bkg}}}
\def\frdstl{f_{R(\dstar\ell)}}
\def\fbgde{F_{\rm bkg}^{\dE}}
\def\fsigde{F_{\rm sig}^{\dE}}
\def\fbgmb{F_{\rm bkg}^{\mb}}
\def\fsigmb{F_{\rm sig}^{\mb}}
\def\rsig{R_{\rm sig}}
\def\rfdst{R_{f \dstar}}
\def\pfl{p_{f \ell}}
\def\pbb{p_{\bbbar}}
\def\pcc{p_{c\overline{c}}}
\def\Rsig{$\rsig$}
\def\rdz{R_{\delz}}
\def\rbg{R_{\rm bkg}}
\def\Rdz{$\rdz$}
\def\tbm{\tau_-}
\def\tbz{\tau_0}
\def\tsig{\tau_{\rm sig}}
\def\tbg{\tau_{\rm bkg}}
\def\BF{{\cal B}}
\def\cosb{\cos\theta_B}
\def\ie{{\it i.e.}}
\def\psig{{{\cal P}_{\rm sig}}}
\def\pbkg{{{\cal P}_{\rm bkg}}}
\def\pasym{{{\cal P}_{\rm asym}}}
\def\rdet{{{R}_{\rm det}}}
\def\rnp{{{R}_{\rm np}}}
\def\rk{{{R}_{\rm k}}}

\newcommand{\Bcp}{B_{CP}}
\newcommand{\Btag}{B_{\rm tag}}
\newcommand{\zcp}{z_{CP}}
\newcommand{\ztag}{z_{\rm tag}}

\newcommand{\Brec}{B_{\rm rec}}
\newcommand{\Basc}{B_{\rm asc}}
\newcommand{\trec}{t_{\rm rec}}
\newcommand{\tasc}{t_{\rm asc}}
\newcommand{\zrec}{z_{\rm rec}}
\newcommand{\zasc}{z_{\rm asc}}
\newcommand{\Rrec}{R_{\rm rec}}
\newcommand{\Rasc}{R_{\rm asc}}
\newcommand{\Srec}{\sigma_{\rm rec}}
\newcommand{\Sasc}{\sigma_{\rm asc}}
\newcommand{\Scp}{\sigma_{CP}}
\newcommand{\Stag}{\sigma_{\rm tag}}
\newcommand{\Sdz}{\sigma_{\Dz}}
\newcommand{\Sdt}{\sigma_{\Dt}}

\newcommand{\tSdt}{\tilde{\sigma}_{\Dt}}
\newcommand{\tSdtm}{\tilde{\sigma}^{\Dt}_{\rm main}}
\newcommand{\tSdtt}{\tilde{\sigma}^{\Dt}_{\rm tail}}

\newcommand{\Sdtm}{\sigma^{\Dt}_{\rm main}}
\newcommand{\Mdtm}{\mu^{\Dt}_{\rm main}}
\newcommand{\Sdtt}{\sigma^{\Dt}_{\rm tail}}
\newcommand{\Mdtt}{\mu^{\Dt}_{\rm tail}}

\newcommand{\tScp}{\tilde{\sigma}_{CP}}
\newcommand{\tStag}{\tilde{\sigma}_{\rm tag}}
\newcommand{\tSdz}{\tilde{\sigma}_{\Dz}}
\newcommand{\tSdzm}{\tilde{\sigma}^{\Dz}_{\rm main}}
\newcommand{\tSdzt}{\tilde{\sigma}^{\Dz}_{\rm tail}}

\newcommand{\Sdzm}{\sigma^{\Dz}_{\rm main}}
\newcommand{\Mdzm}{\mu^{\Dz}_{\rm main}}
\newcommand{\Sdzt}{\sigma^{\Dz}_{\rm tail}}
\newcommand{\Mdzt}{\mu^{\Dz}_{\rm tail}}

\newcommand{\srec}{s_{\rm rec}}
\newcommand{\sasc}{s_{\rm asc}}

\newcommand{\sk}{\sigma_{\rm k}}
\newcommand{\snpm}{s_{\rm main}^{\rm NP}}
\newcommand{\snpt}{s_{\rm tail}^{\rm NP}}
\newcommand{\smain}{s_{\rm main}}
\newcommand{\stail}{s_{\rm tail}}
\newcommand{\ftail}{f_{\rm tail}}

\newcommand{\bra}[1]{\langle#1|}
\newcommand{\ket}[1]{|#1\rangle}
\newcommand{\braket}[2]{\langle#1|#2\rangle}
\newcommand{\rexp}[1]{{\rm e}^{#1}}
\newcommand{\ri}{{\rm i}}
\newcommand{\lcp}{{\lambda_{CP}}}
\newcommand{\dedx}{{\rm d}E/{\rm d}x}
\newcommand{\sinbb}{{\sin2\phi_1}}

\newcommand{\bcdot}{\!\cdot\!}

\def\dzp{\delz^\prime}
\def\fdelbg{f_\delta^{\rm bkg}}
\def\mudelbg{\mu_\delta^{\rm bkg}}
\def\mutaubg{\mu_\tau^{\rm bkg}}

\def\pol{p_{\rm ol}}
\def\fol{f_{\rm ol}}
\def\sigol{\sigma_{\rm ol}}

\def\dM{{\Delta m_d}}

\title{\quad\\[0.5cm] An Improved Measurement of Mixing-induced
{\boldmath $CP$} Violation in the Neutral {\boldmath $B$} Meson System}
%
\affiliation{Aomori University, Aomori}
\affiliation{Budker Institute of Nuclear Physics, Novosibirsk}
\affiliation{Chiba University, Chiba}
\affiliation{Chuo University, Tokyo}
\affiliation{University of Cincinnati, Cincinnati OH}
\affiliation{University of Frankfurt, Frankfurt}
\affiliation{Gyeongsang National University, Chinju}
\affiliation{University of Hawaii, Honolulu HI}
\affiliation{High Energy Accelerator Research Organization (KEK), Tsukuba}
\affiliation{Hiroshima Institute of Technology, Hiroshima}
\affiliation{Institute of High Energy Physics, Chinese Academy of Sciences, Beijing}
\affiliation{Institute of High Energy Physics, Vienna}
\affiliation{Institute for Theoretical and Experimental Physics, Moscow}
\affiliation{J. Stefan Institute, Ljubljana}
\affiliation{Kanagawa University, Yokohama}
\affiliation{Korea University, Seoul}
\affiliation{Kyoto University, Kyoto}
\affiliation{Kyungpook National University, Taegu}
\affiliation{Institut de Physique des Hautes \'Energies, Universit\'e de Lausanne, Lausanne}
\affiliation{University of Ljubljana, Ljubljana}
\affiliation{University of Maribor, Maribor}
\affiliation{University of Melbourne, Victoria}
\affiliation{Nagoya University, Nagoya}
\affiliation{Nara Women's University, Nara}
\affiliation{National Kaohsiung Normal University, Kaohsiung}
\affiliation{National Lien-Ho Institute of Technology, Miao Li}
\affiliation{National Taiwan University, Taipei}
\affiliation{H. Niewodniczanski Institute of Nuclear Physics, Krakow}
\affiliation{Nihon Dental College, Niigata}
\affiliation{Niigata University, Niigata}
\affiliation{Osaka City University, Osaka}
\affiliation{Osaka University, Osaka}
\affiliation{Panjab University, Chandigarh}
\affiliation{Peking University, Beijing}
\affiliation{Princeton University, Princeton NJ}
\affiliation{RIKEN BNL Research Center, Brookhaven NY}
\affiliation{Saga University, Saga}
\affiliation{University of Science and Technology of China, Hefei}
\affiliation{Seoul National University, Seoul}
\affiliation{Sungkyunkwan University, Suwon}
\affiliation{University of Sydney, Sydney NSW}
\affiliation{Tata Institute of Fundamental Research, Bombay}
\affiliation{Toho University, Funabashi}
\affiliation{Tohoku Gakuin University, Tagajo}
\affiliation{Tohoku University, Sendai}
\affiliation{University of Tokyo, Tokyo}
\affiliation{Tokyo Institute of Technology, Tokyo}
\affiliation{Tokyo Metropolitan University, Tokyo}
\affiliation{Tokyo University of Agriculture and Technology, Tokyo}
\affiliation{Toyama National College of Maritime Technology, Toyama}
\affiliation{University of Tsukuba, Tsukuba}
\affiliation{Utkal University, Bhubaneswer}
\affiliation{Virginia Polytechnic Institute and State University, Blacksburg VA}
\affiliation{Yokkaichi University, Yokkaichi}
\affiliation{Yonsei University, Seoul}
  \author{K.~Abe}\affiliation{High Energy Accelerator Research Organization (KEK), Tsukuba} 
  \author{K.~Abe}\affiliation{Tohoku Gakuin University, Tagajo} 
  \author{N.~Abe}\affiliation{Tokyo Institute of Technology, Tokyo} 
  \author{R.~Abe}\affiliation{Niigata University, Niigata} 
  \author{T.~Abe}\affiliation{Tohoku University, Sendai} 
  \author{I.~Adachi}\affiliation{High Energy Accelerator Research Organization (KEK), Tsukuba} 
  \author{Byoung~Sup~Ahn}\affiliation{Korea University, Seoul} 
  \author{H.~Aihara}\affiliation{University of Tokyo, Tokyo} 
  \author{M.~Akatsu}\affiliation{Nagoya University, Nagoya} 
  \author{M.~Asai}\affiliation{Hiroshima Institute of Technology, Hiroshima} 
% ...
\collaboration{The Belle Collaboration}

\date{\today}

\begin{abstract}
We present an improved measurement of the standard model $CP$ violation
parameter $\sinbb$ (also known as $\sin2\beta$) based on a sample
of $85\times10^6$
$B\overline{B}$ pairs collected at the $\Upsilon(4S)$ resonance
with the Belle detector at the KEKB asymmetric-energy $e^+e^-$ collider.
One neutral $B$ meson is reconstructed in a $\jpsi\ks$, $\psi(2S)\ks$,
$\chi_{c1}\ks$, $\eta_c\ks$, $\jpsi K^{*0}$, or $\jpsi\kl$ $CP$-eigenstate
decay channel and the flavor of accompanying $B$ meson is identified from its
decay products.  From the asymmetry in the distribution of the time intervals
between the two $B$ meson decay points, we obtain
$\sinbb=0.719\pm0.074\mbox{(stat)}\pm0.035\mbox{(syst)}$.
We also report measurements of $CP$ violation parameters for
the related $\bz\to \jpsi \pi^0$ decay mode and the
penguin-dominated processes $\bz\to \eta' \ks$,
$\phi \ks$ and $K^+K^- \ks$.

\end{abstract}


\maketitle

In the Standard Model (SM), $CP$ violation arises from an
irreducible complex phase in the weak interaction quark-mixing matrix
(CKM matrix)~\cite{bib:ckm}.
In particular, the SM predicts a $CP$-violating asymmetry
in the time-dependent rates for $\bz$ and $\bb$ decays to a common
$CP$ eigenstate, $f_{CP}$,
with negligible corrections from strong interactions\cite{bib:sanda}:
\be
A(t) \equiv \frac{\Gamma(\bb\to f_{CP}) - \Gamma(\bz\to f_{CP})}
{\Gamma(\bb\to f_{CP}) + \Gamma(\bz\to f_{CP})} = -\xi_f \sinbb \sin(\dM t),
\ee
where $\Gamma(\bz,\bb \to f_{CP})$ is the decay rate for a $\bz$ or $\bb$
to $f_{CP}$ dominated by a $b\to c\overline{c}s$ transition
at a proper time $t$ after production, $\xi_f$ is the $CP$ eigenvalue of
$f_{CP}$, $\dM$ is the mass difference between the two $\bz$ mass eigenstates,
and $\phi_1$ is one of the three interior angles of the CKM unitarity triangle,
defined as $\phi_1 \equiv \pi-\arg(-V_{tb}^*V_{td}/-V_{cb}^*V_{cd})$.
Non-zero values for $\sin 2\phi_1$ were reported by the Belle
and BaBar groups\cite{bib:cpv,bib:babar}.

Belle's published measurement of
$\sin 2\phi_1$ is based on a 29.1~fb$^{-1}$ data sample
containing $31.3\times 10^{6}$ $B\overline{B}$ pairs
produced at the $\Upsilon(4S)$ resonance.
In this paper, we report an improved measurement that uses
$85\times 10^6$ $B\overline{B}$ pairs (78~fb$^{-1}$).
The data were collected with the Belle detector~\cite{bib:belle}
at the KEKB  asymmetric collider~\cite{bib:kekb}, 
which collides 8.0~GeV $e^-$ on
3.5~GeV $e^+$ at a small ($\pm 11$~mrad) crossing angle.
We use events where one of the $B$ mesons decays to $f_{CP}$ at time $t_{CP}$,
and the other decays
to a self-tagging state, $f_{\rm tag}$, {\it i.e.}, a final state that
distinguishes $\bz$ and $\bb$, at time $t_{\rm tag}$.
The $CP$ violation manifests itself as an asymmetry $A(\Dt)$,
where $\Dt$ is the proper time interval
between the two decays: $\Dt \equiv t_{CP}-t_{\rm tag}$.
At KEKB, the $\Upsilon(4S)$ resonance is produced with a
boost of $\beta\gamma=0.425$ nearly along the electron beam direction ($z$
direction), and
$\Dt$ can be determined as $\Dt \simeq \Dz/(\beta\gamma)c$,
where $\Dz$ is the $z$ distance
between the $f_{CP}$ and $f_{\rm tag}$ decay vertices, $\Dz \equiv \zcp-\ztag$.
The $\Dz$ average value is approximately 200 $\mu$m.

The Belle detector~\cite{bib:belle} is a large-solid-angle
spectrometer that
consists of a silicon vertex detector (SVD),
a central drift chamber (CDC), an array of
aerogel threshold \v{C}erenkov counters (ACC),
time-of-flight
scintillation counters (TOF), and an electromagnetic calorimeter
comprised of CsI(Tl) crystals (ECL)  located inside
a superconducting solenoid coil that provides a 1.5~T
magnetic field.  An iron flux-return located outside of
the coil is instrumented to detect $K_L^0$ mesons and to identify
muons (KLM).

\begin{figure}[htbp]
\begin{center}
    \includegraphics[width=0.45\textwidth,clip]{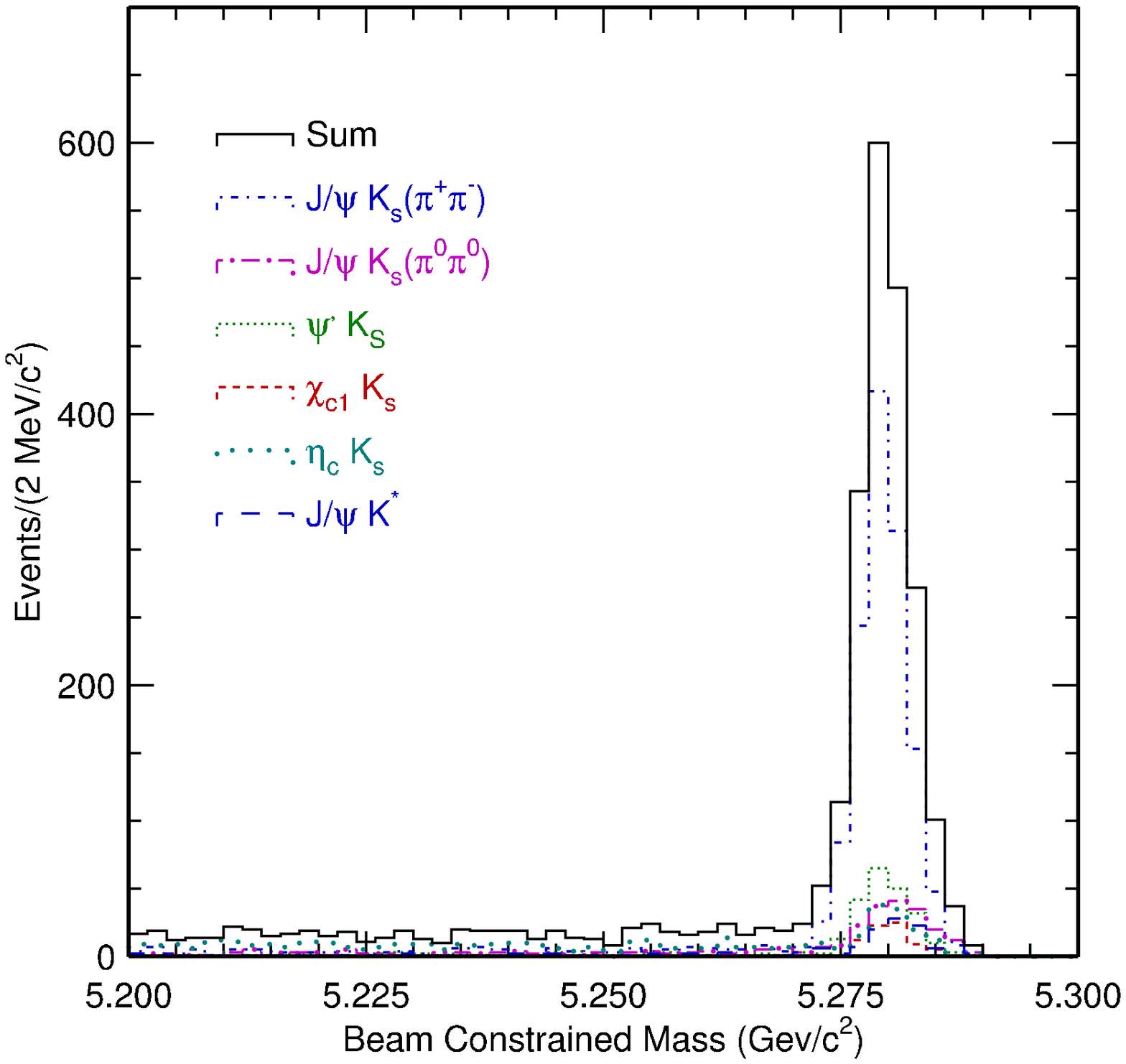}
    \includegraphics[width=0.49\textwidth,clip]{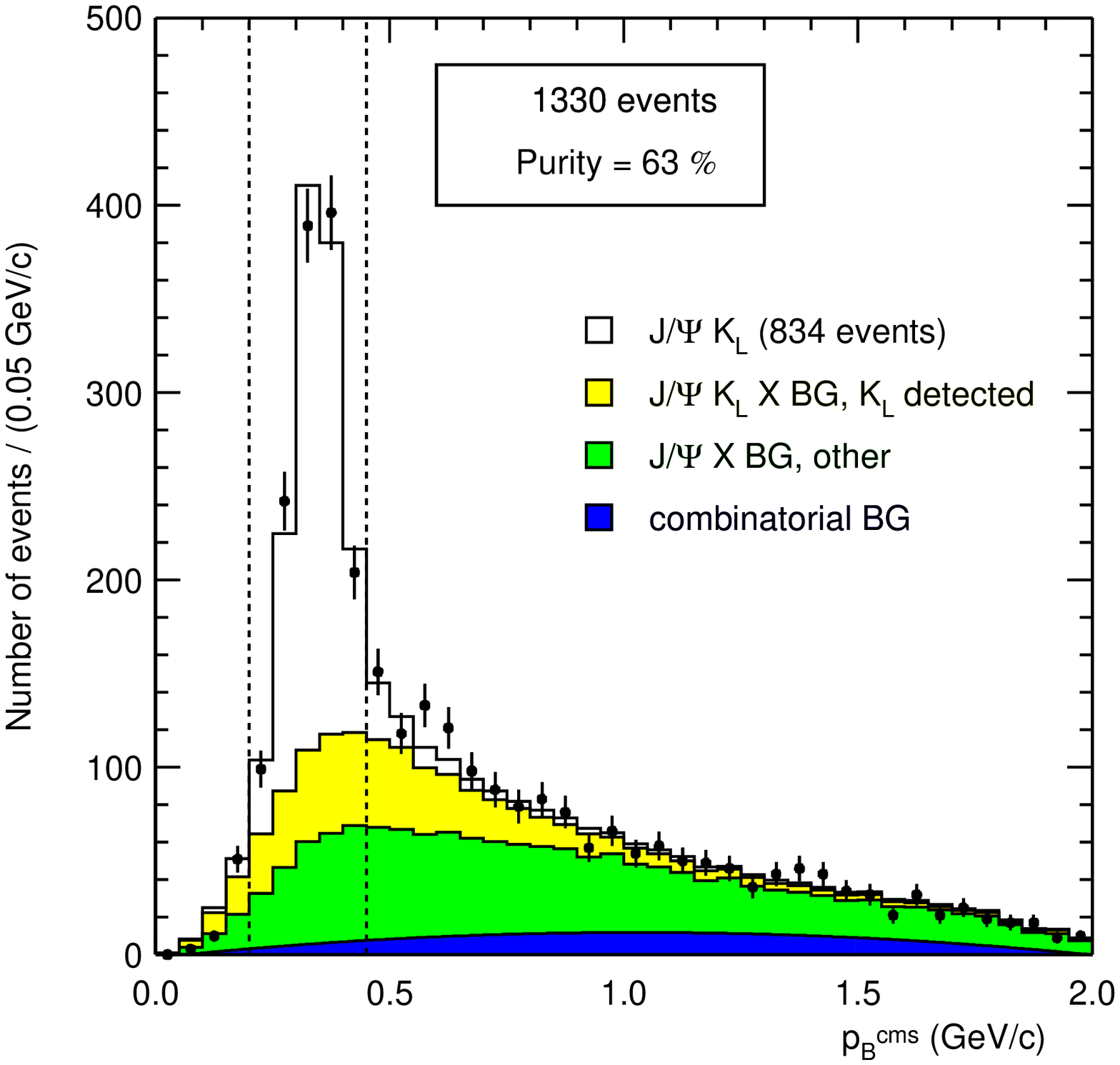}
  \end{center}
\caption{The beam-energy constrained mass
distribution for all decay modes
combined other than $\jpsi\kl$ (left).
The $p_B^{\rm cms}$ distribution for
$\bz\to\jpsi\kl$ candidates with the results of the fit (right).}
\label{fig:mbc}
\end{figure}

We reconstruct $\bz$ decays to the following $CP$ 
eigenstates~\cite{footnote:cc}:
$\jpsi\ks$, $\psi(2S)\ks$, $\chi_{c1}\ks$, $\eta_c\ks$ for $\xi_f = -1$ and
$\jpsi\kl$ for $\xi_f = +1$.  We also use $\bz\to\jpsi K^{*0}$ decays where
$K^{*0} \to \ks\piz$.  Here the final state is a mixture of even and odd
$CP$, depending on the relative orbital angular momentum of the $\jpsi$
and $K^{*0}$.  We find that the final state is primarily $\xi_f = +1$; the
$\xi_f=-1$ fraction is
$0.19\pm0.02\mbox{(stat)}\pm0.03\mbox{(syst)}$\cite{bib:itoh}.
For reconstructed $B\to f_{CP}$ candidates other than $\jpsi\kl$,
we identify $B$ decays using the
energy difference $\dE\equiv E_B^{\rm cms}-E_{\rm beam}^{\rm cms}$ and the
beam-energy constrained mass
$\mb \equiv \sqrt{(E_{\rm beam}^{\rm cms})^2-(p_B^{\rm cms})^2}$, where
$E_{\rm beam}^{\rm cms}$ is the beam energy in the 
center-of-mass system (cms), and
$E_B^{\rm cms}$ and $p_B^{\rm cms}$ are 
the cms energy and momentum of the reconstructed $B$ candidate,
respectively.
Figure \ref{fig:mbc} (left) shows the $\mb$ distributions for all
$\bz$ candidates except for $\bz\to\jpsi\kl$ that 
have $\dE$ values in the signal region.
\begin{table}[hbtp]
\caption{The numbers of reconstructed $B \to f_{CP}$
candidates before flavor tagging and vertex reconstruction
($N_{\rm rec}$),
the numbers of events used for the $\sin 2\phi_1$ determination
($N_{\rm ev}$), and the
estimated signal purity in the signal region for each $f_{CP}$ mode.}
\begin{ruledtabular}
\begin{tabular}{lcrrr}
Mode & $\xi_f$ & $N_{\rm rec}$ & $N_{\rm ev}$ & Purity \\
\hline
$\jpsi(\ell^+\ell^-)\ks(\pi^+\pi^-)$ & $-1$     & 1285 & 1116 & 0.98 \\
$\jpsi(\ell^+\ell^-)\ks(\pi^0\pi^0)$ & $-1$      & 188 & 162   & 0.82 \\
$\psi(2S)(\ell^+\ell^-)\ks(\pi^+\pi^-)$ &$-1$    &  91 & 76    & 0.96 \\
$\psi(2S)(\jpsi\pi^+\pi^-)\ks(\pi^+\pi^-)$ &$-1$ & 112 & 96    & 0.91 \\
$\chi_{c1}(\jpsi\gamma)\ks(\pi^+\pi^-)$ &$-1$    &  77 & 67    & 0.96 \\
$\eta_c(\ks K^-\pi^+)\ks(\pi^+\pi^-)$ &$-1$      &  72 & 63    & 0.65 \\
$\eta_c(K^+K^-\pi^0)\ks(\pi^+\pi^-)$ &$-1$       &  49 & 44    & 0.72 \\
$\eta_c(p\overline{p})\ks(\pi^+\pi^-)$ &$-1$     &  21 & 15    & 0.94 \\
\hline
\multicolumn{2}{l}{All with $\xi_f = -1$}        &1895 & 1639  & 0.94 \\
\hline
$\jpsi(\ell^+\ell^-) K^{*0}(\ks\pi^0)$& $-1$(19\%)/+1(81\%)& 101 & 89    & 0.92 \\
\hline
$\jpsi(\ell^+\ell^-) \kl$ &+1                  & 1330 &1230  & 0.63 \\
\hline \hline
\multicolumn{2}{l}{All}                        & 3326 &2958  & 0.81 \\
\end{tabular}
\end{ruledtabular}
\label{tab:number}
\end{table}
Table~\ref{tab:number} lists the
numbers of observed candidates ($N_{\rm rec}$).

Candidate $\bz\to\jpsi\kl$ decays are selected by requiring
ECL and/or KLM
hit patterns that are consistent with the presence of a shower
induced by a neutral hadron.
The centroid of the shower is required to be
in a $45^\circ$ cone centered on the $\kl$ direction that is inferred from
two-body decay kinematics and the measured four-momentum of the $\jpsi$.
Figure \ref{fig:mbc} (right) shows the $p_B^{\rm cms}$ distribution,
calculated with the $\bz\to\jpsi\kl$ two-body decay hypothesis.  
The histograms
are the results of a fit to the signal and background distributions.
There are 1330 entries in total in the
$0.20\le p_B^{\rm cms}\le0.45$~GeV/$c$ signal region;
the fit indicates a signal purity of 63\%. The reconstruction and
selection criteria for all of $f_{CP}$ channels
used in the measurement are described in 
more detail elsewhere~\cite{bib:cpv}.

Leptons, charged pions, kaons, and $\Lambda$ baryons that are not associated
with a reconstructed $CP$ eigenstate decay are used to identify the $b$-flavor
of the accompanying $B$ meson:
high momentum leptons from $b\to c\ell^-\overline{\nu}$;
lower momentum leptons from $c\to s\ell^+\nu$;
charged kaons and $\Lambda$ baryons from $b\to c \to s$;
fast pions from $\bz\to D^{(*)-}$($\pi^+,\rho^+,a_1^+$, etc.); and
slow pions from $D^{*-}\to \overline{D}{}^0\pi^-$.  Based on the
measured properties of these tracks, two parameters, $q$ and $r$, are
 assigned to an event.
The first, $q$, has the discrete values $q=\pm1$ that is $+1~(-1)$
when $\Btag$ is likely to be a $\bz$~($\bb$), and the parameter $r$ is an
event-by-event Monte-Carlo-determined
flavor-tagging dilution factor
that ranges from $r=0$ for no flavor
discrimination to $r=1$ for an unambiguous flavor assignment.  It is used only
to sort data into six intervals of $r$, according to flavor purity;
the wrong-tag probabilities, $w_l~(l=1,6)$, that are used in
the final fit are determined directly from data.  Samples of
$B^0$ decays to exclusively reconstructed self-tagged channels
are utilized to obtain $w_l$ using time-dependent $\bz$-$\bb$ mixing
oscillation:
$(N_{\rm OF}-N_{\rm SF})/(N_{\rm OF}+N_{\rm SF}) = (1-2w_l)\cos(\dM\Dt)$,
where $N_{\rm OF}$ and $N_{\rm SF}$ are the numbers of opposite
and same flavor events.
The event fractions and wrong tag fractions are summarized in
Table~\ref{tab:wtag}.
The total effective tagging efficiency is determined to be
$\sum_{l=1}^6f_l(1-2w_l)^2 = 0.288\pm0.006$,
where $f_l$ is the event fraction for each $r$ interval.
\begin{table}[hbtp]
\caption{The event fractions ($\epsilon_l$) and wrong tag
fractions ($w_l$) fo each $r$ interval. The errors include
both statistical and systematic uncertainties. The
event fractions are obtained from the $J/\psi\ks$ simulation.}
\begin{ruledtabular}
\begin{tabular}{cccl}
$l$ & $r$ & $\epsilon_l$ &\multicolumn{1}{c}{$w_l$} \\
\hline
1 & 0.000 $-$ 0.250 & 0.399 & $0.458\pm0.006$ \\
2 & 0.250 $-$ 0.500 & 0.146 & $0.336\pm0.009$ \\
3 & 0.500 $-$ 0.625 & 0.104 & $0.229~^{+0.010}_{-0.011}$ \\
4 & 0.625 $-$ 0.750 & 0.122 & $0.159\pm0.009$ \\
5 & 0.750 $-$ 0.875 & 0.094 & $0.111\pm0.009$ \\
6 & 0.875 $-$ 1.000 & 0.137 & $0.020~^{+0.007}_{-0.006}$ \\
\end{tabular}
\end{ruledtabular}
\label{tab:wtag}
\end{table}

The vertex position for the $f_{CP}$ 
decay is reconstructed using leptons from $\jpsi$
decays or kaons and pions from $\eta_c$
and that for $f_{\rm tag}$ is obtained 
with well reconstructed tracks
that are not assigned to $f_{CP}$.  Tracks that are consistent
with coming from a $\ks\to\pi^+\pi^-$ decay  are not used.
Each vertex position is required to be consistent with
a run-by-run-determined interaction region
profile that is smeared in the $r$-$\phi$ plane by the $B$ meson decay
length.  With these requirements, we are able to determine a vertex
even with a single track;
the fraction of single-track vertices is about 10\%
for $\zcp$ and 22\% for $\ztag$.
The proper-time interval resolution function, $R_{\rm sig}(\Dt)$,
is formed by convolving four components:
the detector resolutions for $\zcp$ and $\ztag$,
the shift in the $\ztag$ vertex position
due to secondary tracks originating from
charmed particle decays, and
smearing due to the kinematic approximation
used to convert $\Dz$ to $\Dt$.
A small component of broad outliers in the $\Dz$ distribution,
caused by mis-reconstruction,  is represented by a Gaussian
function.  We determine
twelve resolution parameters from the data
from fits to the neutral
and charged $B$ meson lifetimes~\cite{bib:lifetime}
and obtain an average $\Dt$ resolution of $\sim 1.43$~ps (rms).
The width of the outlier component
is determined to be $(42^{+5}_{-4})$~ps;
the fractional areas are $(2 \pm 1)\times 10^{-4}$ and
$(2.7\pm0.2)\times 10^{-2}$ for the multiple- and  
single-track cases, respectively.

\begin{figure}[htbp]
\begin{center}
\includegraphics[width=0.6\textwidth,clip]{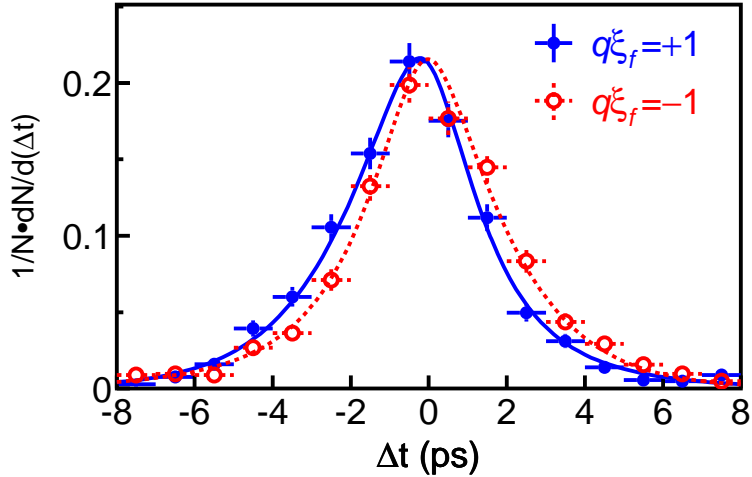}
\end{center}
\caption{$\Dt$ distributions for the events with $q\xi_f=+1$ (solid points)
and $q\xi_f=-1$ (open points).  The results of the global fit with
$\sinbb=0.719$ are shown as solid and dashed curves, respectively.}
\label{fig:cpfit}
\end{figure}

After flavor tagging and vertexing,
we find 1465 events with $q=+1$ flavor tags and 1493 events
with $q=-1$.
Table~\ref{tab:number} lists the numbers of candidates used
for the $\sin 2\phi_1$ determination ($N_{\rm ev}$) and
the estimated signal purity in the signal region for each
$f_{CP}$ mode.
Figure \ref{fig:cpfit} shows the observed $\Dt$ distributions
for the $q\xi_f=+1$ (solid points) 
and $q\xi_f=-1$ (open points) event samples.
The asymmetry between 
the two distributions demonstrates the violation of
$CP$ symmetry.
We determine $\sinbb$ from an unbinned maximum-likelihood fit to the observed
$\Dt$ distributions.  The probability density function (pdf) expected
for the signal distribution is given by
\be
\label{eq:deltat}
{\cal P}_{\rm sig}(\Dt,q,w_l,\xi_f) =
\frac{e^{-|\Dt|/\tau_\bz}}{4\tau_\bz}[1-q\xi_f(1-2w_l)\sinbb\sin(\dM\Dt)],
\ee
where we fix the $\bz$ lifetime ($\tau_\bz$) and mass difference
at their world average values\cite{bib:pdg}.
Each pdf is convolved with the appropriate $R_{\rm sig}(\Dt)$
to determine the likelihood value for each event as a function of $\sinbb$:
\begin{eqnarray}
P_i &=& (1-f_{\rm ol})\int \Bigl[ f_{\rm sig}{\cal P}_{\rm sig}(\Dt',q,w_l,\xi_f)R_{\rm sig}(\Dt-\Dt') \nonumber \\
&& +\; (1-f_{\rm sig}){\cal P}_{\rm bkg}(\Dt')R_{\rm bkg}(\Dt-\Dt')\Bigr] d\Dt'
+ f_{\rm ol}P_{\rm ol}(\Dt),
\end{eqnarray}
where $f_{\rm sig}$ is the signal probability calculated
as a function of $p_B^{\rm cms}$ for $\jpsi\kl$ and of $\dE$ and $\mb$ for
other modes.
${\cal P}_{\rm bkg}(\Dt)$ is the pdf for combinatorial background events,
which is modeled as a sum of exponential and prompt components.  It is
convolved with a sum of two Gaussians, $R_{\rm bkg}$, which is
regarded as a resolution function for the background.
To account for a small number of events that give large $\Delta t$ in
both the signal and background, we introduce the pdf,
$P_{\rm ol}$, and the fractional area, $f_{\rm ol}$,
of the outlier component.
The only free parameter in the final fit is $\sinbb$, which is determined by
maximizing the likelihood function $L=\prod_iP_i$, where the product is over
all events.  The result of the fit is
\[
\sinbb = 0.719\pm 0.074\mbox{(stat)}\pm0.035\mbox{(syst)} .
\]
\begin{figure}[htbp]
\begin{center}
\includegraphics[width=0.8\textwidth,clip]{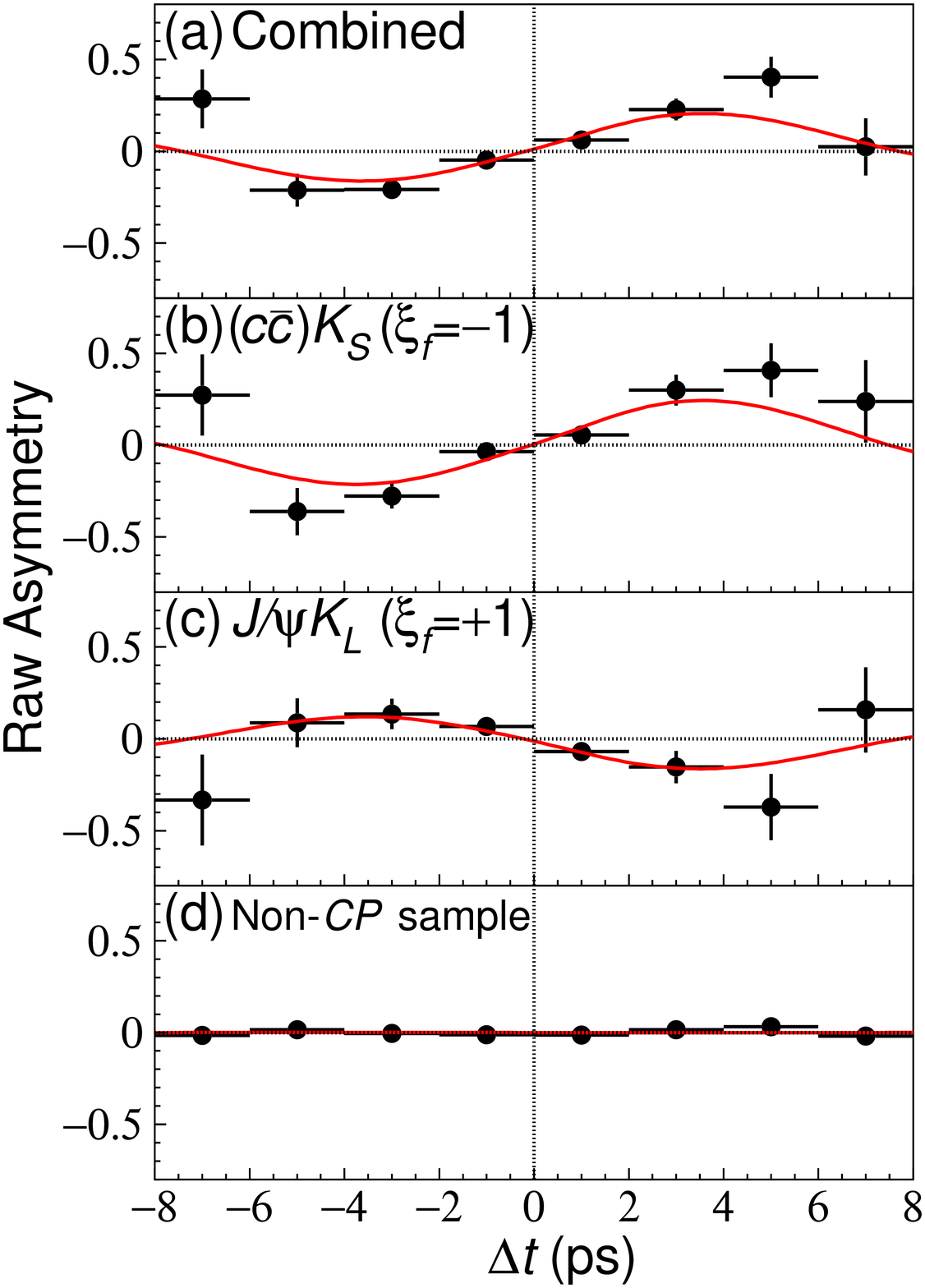}
\end{center}
\caption{(a) The raw asymmetry for all modes combined.
The asymmetry for $\jpsi\kl$ and $\jpsi K^{*0}$ is inverted
to account for the opposite $CP$ eigenvalue.
The corresponding plots for (b) $(c\overline{c})\ks$,
(c) $\jpsi\kl$, and (d) non-$CP$ control samples are also shown.
The curves are the results of the unbinned maximum likelihood fit
applied separately to the individual data samples.}
\label{fig:rawasym}
\end{figure}
%
The systematic error is dominated by uncertainties
in the vertex reconstruction (0.022). Other significant
contributions come from uncertainties in:
$w_l$ (0.015);
the resolution function parameters (0.014);
a possible bias in the $\sin 2\phi_1$ fit (0.011);
and the $\jpsi\kl$ background fraction (0.010).
The errors introduced by uncertainties in $\dM$ and $\tau_\bz$ are
less than 0.010.

A number of checks on the measurement are performed.  Table~\ref{tab:check}
lists the results obtained by applying the same analysis to various subsamples.
\begin{table}[hbtp]
\caption{The numbers of candidate events ($N_{\rm ev}$) and
values of $\sinbb$ for various subsamples (statistical errors only).}
\begin{ruledtabular}
\begin{tabular}{lrc}
Sample & $N_{\rm ev}$ & $\sinbb$ \\
\hline
$\jpsi\ks(\pip\pim)$& 1116                         &  $0.73\pm0.10$ \\
$(c\overline{c})\ks$ except $\jpsi\ks(\pip\pim)$&523& $0.67\pm0.17$ \\
$\jpsi\kl$ &1230                                    & $0.78\pm0.17$ \\
$\jpsi K^{*0}(\ks\pi^0)$& 89                        & $0.04\pm0.63$ \\
\hline
$f_{\rm tag}=\bz$ $(q=+1)$& 1465                   &  $0.65\pm0.12$ \\
$f_{\rm tag}=\bb$ $(q=-1)$& 1493                   &  $0.77\pm0.09$ \\
\hline
$0 < r \le 0.5$ &   1600 & $1.26\pm0.36$ \\
$0.5 < r \le 0.75$&  658 & $0.62\pm0.15$ \\
$0.75 < r \le 1$&    700 & $0.72\pm0.09$\\
\hline \hline
All & 2958                                          & $0.72\pm0.07$ \\
\end{tabular}
\end{ruledtabular}
\label{tab:check}
\end{table}
All values are statistically consistent with each other.
Figure~\ref{fig:rawasym}(a), (b), and (c) show the raw asymmetries and the
fit results for all modes combined,
$(c\overline{c}) \ks$, and $\jpsi\kl$, respectively.
A fit to the non-$CP$ eigenstate self-tagged modes $\bz\to D^{(*)-}\pi^+$,
$D^{*-}\rho^+$, $\jpsi K^{*0}(K^+\pi^-)$, and $D^{*-}\ell^+\nu$,
where no asymmetry is expected,
yields $0.004\pm0.015$(stat). Figure~\ref{fig:rawasym}(d) shows
the raw asymmetry for these non-$CP$ control samples.

The signal pdf for a neutral $B$ meson decaying into a $CP$ eigenstate
(Eq.~(\ref{eq:deltat}))
can be expressed in a more general form as
\begin{eqnarray}
\label{eq:deltat_general}
{\cal P}_{\rm sig}(\Delta t,q,w_l,\xi_f) =
 \frac{ e^{-|\Delta t|/\tau_{B^0}} }{4\tau_{B^0}}
\Bigl\{ 1 + q(1-2w_l)
\Bigl[ \frac{2|\lambda|(-\xi_f)a_{CP}}{|\lambda|^2+1}\sin(\Delta m_d\Delta t)
\nonumber \\
    + \frac{|\lambda|^2 -1}{|\lambda|^2+1} \cos(\Delta m_d\Delta t) \Bigr]
\Bigr\},
\end{eqnarray}
where $\lambda$ is a complex parameter
that depends on both
$\bz$-$\bb$ mixing and on the amplitudes for $B^0$ and $\bzb$ decay
to a $CP$ eigenstate.
The parameter $a_{CP}$ in
the coefficient of $\sin(\Delta m_d\Delta t)$ is given by
$a_{CP} = -\xi_f Im\lambda/|\lambda|$
and is equal to $\sin 2 \phi_1$ in the SM.
The presence of the cosine term
($|\lambda| \neq 1$) would indicate direct $CP$ violation;
the value for $\sin 2\phi_1$ reported above is determined 
with the assumption 
$|\lambda| = 1$, as expected in the SM.
In order to test this assumption,
we also performed a fit using the above expression with
$a_{CP}$
and $|\lambda|$ as free parameters,
keeping everything else the same.
We obtain
\[
  |\lambda| = 0.950 \pm 0.049\mbox{(stat)} \pm 0.026\mbox{(syst)}
\]
and $a_{CP} = 0.720 \pm 0.074\mbox{(stat)}$
for all $CP$ modes combined, where the sources of the systematic
error for $|\lambda|$ are the same as those for $\sin 2\phi_1$.
This result confirms the assumption used in our analysis.

Finally, we also report on the time-dependent $CP$ asymmetries
in the $\bz\to\eta'\ks$,
$\bz\to\phi\ks$ and $\bz\to K^+K^-\ks$ decays that are
dominated by the $b \to s$ penguin diagrams,
and in the $\bz\to \jpsi \pi^0$ decay governed by
the Cabibbo-suppressed $b \to c\overline{c}d$ transition.
All measurements are based on $85\times10^6$ $B\overline{B}$ pairs.
The reconstruction methods for these modes are described
elsewhere~\cite{bib:othercp,bib:kkks}.
The signal yields are summarized in Table~\ref{tab:othercp}.
The flavor-tagging, vertexing
and fitting techniques are the same as those described above.
The decay rate has a time dependence given by
\be
{\cal P}_{\rm sig}(\Dt,q,w_l) = \frac{e^{-|\Dt|/\tau_\bz}}{4\tau_\bz}\Bigl\{1+q(1-2w_l)[{\cal S}\sin(\dM\Dt)+{\cal A}\cos(\dM\Dt)]\Bigr\}.
\ee
Table~\ref{tab:othercp} summarizes the results of
the fits to the $\Dt$ distributions.
In the table, we show the values of $-\xi_f{\cal S}$
denoted by ``$\sin2\phi_1$'', since
it should be equal to $\sin 2\phi_1$ in
the limit that new physics in the penguin loop does not
contribute additional $CP$-violating phases.
\begin{table}[hbtp]
\caption{Results of the $CP$ asymmetry parameter measurements for
$\bz\to\eta'\ks$, $\bz\to\phi\ks$, $\bz\to K^+K^-\ks$, and
$\bz\to \jpsi \pi^0$ decays.
``$\sin2\phi_1$'' is defined as $-\xi_f{\cal S}$ and should be equal to
$\sin2\phi_1$ in the limit of no contributions from new physics
in the penguin loop.
For ``$\sin2\phi_1$'' and ${\cal A}$,
the first errors are statistical and the
second errors are systematic. The third errors for
the $K^+K^-\ks$ mode arise from the uncertainty in the
fraction of the $CP$-odd component~\cite{bib:kkks}.}
\begin{ruledtabular}
\begin{tabular}{lcrcll}
Mode         & $\xi_f$ & $N_{\rm ev}$
                             &Purity (\%)
                                   &\multicolumn{1}{c}{``$\sin2\phi_1$''}
                                   &\multicolumn{1}{c}{${\cal A}$}\\
\hline
$\eta'\ks$   & $-1$    & 299 & 49  & $+0.76\pm0.36^{+0.05}_{-0.06}$ 
                                                           & $+0.26\pm0.22\pm0.03$ \\
$\phi\ks$    & $-1$    &  53 & 67  & $-0.73\pm0.64\pm0.18$ & $-0.56\pm0.41\pm0.12$ \\
$K^+K^-\ks$  & $-1(3\%)/+1(97\%)$
                       & 191 & 49  & $+0.52\pm0.46\pm0.11^{+0.27}_{-0.03}$
                                                           & $-0.42\pm0.36\pm0.09^{+0.03}_{-0.22}$ \\
$\jpsi\pi^0$ & $+1$    &  57 & 86  & $+0.93\pm0.49\pm0.08$ & $-0.25\pm0.39\pm0.06$ \\
\end{tabular}
\end{ruledtabular}
\label{tab:othercp}
\end{table}

We wish to thank the KEKB accelerator group for the excellent
operation of the KEKB accelerator.
We acknowledge support from the Ministry of Education,
Culture, Sports, Science, and Technology of Japan
and the Japan Society for the Promotion of Science;
the Australian Research Council
and the Australian Department of Industry, Science and Resources;
the National Science Foundation of China under contract No.~10175071;
the Department of Science and Technology of India;
the BK21 program of the Ministry of Education of Korea
and the CHEP SRC program of the Korea Science and Engineering Foundation;
the Polish State Committee for Scientific Research
under contract No.~2P03B 17017;
the Ministry of Science and Technology of the Russian Federation;
the Ministry of Education, Science and Sport of the Republic of Slovenia;
the National Science Council and the Ministry of Education of Taiwan;
and the U.S.\ Department of Energy.



\begin{thebibliography}{99}
\bibitem{footnote:cc}
Throughout this paper, when a decay mode is quoted,
the inclusion of the charge conjugation mode is implied.
\end{thebibliography}
\end{document}